\documentclass[english,11pt]{article}
          \usepackage{babel}
          \usepackage[T1]{fontenc}
          \usepackage[latin2]{inputenc}
          \usepackage{pslatex}
\begin{document}

\title{\bf Curious Variables Experiment (CURVE). \\
CCD Photometry of V419 Lyr in its 2006 July Superoutburst.}
\author{A. R~u~t~k~o~w~s~k~i$^1$, ~~A. ~O~l~e~c~h$^1$,  ~~K. ~M~u~l~a~r~c~z~y~k~$^2$,
~~D. ~B~o~y~d$^3$, 
\\~~ R. ~K~o~f~f$^4$ ~~and~~ M. ~W~i~\'s~n~i~e~w~s~k~i$^1$}

\date{$^1$ Nicolaus Copernicus Astronomical Center,
Polish Academy of Sciences,
ul.~Bartycka~18, 00-716~Warszawa, Poland,\\
{\tt e-mail: (rudy,olech,mwisniew)@camk.edu.pl}\\
~\\
$^2$ Warsaw University Observatory, Al. Ujazdowskie 4, 00-476 Warszawa,
Poland, \\ {\tt e-mail: kmularcz@astrouw.edu.pl}\\
~\\
$^3$ British Astronomical Association, Variable Star Section, \\
West Challow OX12 9TX, England\\
{\tt e-mail: drsboyd@dsl.pipex.com}\\
~\\
$^4$ Antelope Hills Observatory, 980 Antelope Drive West, \\
Bennett, CO 80102, USA \\
{\tt e-mail: bob@antelopehillsobservatory.org}
}

\maketitle

\begin{abstract}

We report extensive photometry of the dwarf nova V419 Lyr throughout its
2006 July superoutburst till quiescence. The superoutburst with
amplitude of $\sim 3.5$ magnitude lasted at least 15 days and was
characterized by the presence of clear superhumps with a mean period of
$P_{\rm sh}=0.089985(58)$ days ($129.58\pm0.08$ min). According to the
Stolz-Schoembs relation, this indicates that the orbital period of the
binary should be around 0.086 days i.e. within the period gap.

During the superoutburst the superhump period was decreasing with the
rate of $\dot{P}/P_{sh}=-24.8(2.2)\times10^{-5}$, which is one of the
highest values ever observed in SU UMa systems. At the end of the plateau
phase, the superhump period stabilized at a value of 0.08983(8) days.

The superhump amplitude decreased from 0.3 mag at the beginning
of the superoutburst to 0.1 mag at its end. In the case of V419 Lyr we
have not observed clear secondary humps, which seems to be typical
for long period systems.

\end{abstract}

\maketitle

\section{Introduction}

Dwarf novae -- a subclass of Cataclysmic Variable stars -- are quite
well studied interacting binary systems composed of late-type red dwarf
secondary and white dwarf primary stars (Warner 1995, Hellier 2001).
Matter transferred from the red dwarf forms an accretion disc around the
white dwarf. Although in the last decade significant progress has been
made in explaining the behaviour of dwarf novae light curves, some
physical processes ongoing in these systems are still not fully
understood (see for example Smak 2000, Schreiber and Lasota 2007). In
particular, the thermal-tidal instability model of Osaki (1996, 2005)
describing the phenomenon of superoutbursts and superhumps may be tested
by examination of SU UMa-type dwarf novae light curves. Additionally,
objects near and inside the so called period gap are very important from
an evolutionary point of view. Those systems give us an unprecedented
opportunity to study the evolution of dwarf novae.

V419 Lyr is a poorly studied cataclysmic variable discovered by
Kurochkin (1990) and originally classified as a Z Cam-type dwarf nova.
Later, Nogami et al. (1998) caught this object in outburst and found
superhumps in its light curve. Detection of superhumps together with
characteristic properties of the outburst allowed them to classify V419
Lyr as a SU UMa-type dwarf nova, but short coverage of the eruption did
not allow accurate determination of the superhump period.
Nevertheless, there was a strong suggestion that V419 Lyr has one of
the longest orbital periods known among SU UMa variables. 

This object has been monitored at various photometric bands by the
Variable Star Network (VSNET) (see for example Kato et al. 2004a). The
observations from that program enabled a tentative determination of the
supercycle period to be about $\sim 340$ days (Katysheva and Pavlenko
2003). Moreover, Morales-Rueda and Marsh (2002) obtained a spectrum of
V419 Lyr during outburst showing a relatively broad absorption feature
around 430-440 nm.

In this work we present an analysis of photometric data collected during
the 2006 July superoutburst of V419 Lyr. The data are much richer than
previous studies and provide us with an opportunity to determine
parameters describing this system more precisely. 
 
\section{Observations and data reduction}

The CURious Variable Experiment (CURVE) team (see for example Olech et
al. 2004, 2006), alerted by the VSNET mailing list, found V419 Lyr in a
very bright state on 2006 July 17/18. Subsequently the object was
monitored on 13 consecutive nights (with a gap on July 24/25) until its
return to quiescence on August 2/3. The observations were performed
using a 0.6-m Cassegrain telescope equipped with a Tektronix TK512CB
back-illuminated CCD camera. The image scale was $0''.76/$pixel
providing a $6'.5 \times 6'.5$ field of view (Udalski and Pych, 1992)

Observations were made unfiltered for two reasons. First, due to lack of
an autoguiding system, we wished to keep exposures short in order to
minimize guiding errors. Second, because our main goal was analysis of
the temporal behaviour of the light curve, the use of filters might
cause the object to be too faint to observe in quiescence. Exposure
times were 90 seconds during the bright state and 100-150 seconds at
minimum light.

All CURVE team data reductions were performed using a standard procedure
based on IRAF\footnote{ IRAF is distributed by the National Optical
Astronomy Observatory, which is operated by the Association of
Universities for Research in Astronomy, Inc., under cooperative
agreement with the National Science Foundation.} package and the profile
photometry has been derived using the DAOphotII package (Stetson 1987).

During preliminary analysis of the data we found the AAVSO archive
containing several CCD observations of the same superoutburst of V419
Lyr made by observers from England (D.B.) and the United States (R.K.).
Therefore we decided to combine our data to obtain better results.

D.B. used a 0.35-m Meade Schmidt-Cassegrain telescope with a Starlight
Xpress SXV-H9 CCD camera. Data were taken unfiltered. Exposures were 40,
50 or 60 sec depending on conditions. The AIP4WIN package (Berry and
Burnell 2000) was used to dark-subtract and flat-field all images before
measuring them using aperture photometry. The magnitude of V419 Lyr was
determined by differential photometry with respect to an ensemble of two
nearby comparison stars.

R.K. used 0.25-m Meade LX-200 Schmidt-Cassegrain telescope equipped with
an Apogee AP-47 CCD camera and clear filter characterized by IR block at
700 nm. MPO Canopus software (Warner 2007) was used for differential
photometry of the variable with an average of four comparison stars.

Table 1 presents a journal of our CCD observations of V419 Lyr. In
total, we observed the star for more than 48 hours on 14 nights and
collected 3179 exposures. 

\begin{table}[h!]
\caption{Observational journal for the V419 Lyr campaign. O. Start and 
O. End correspond (for particular nights) to times for first and 
last CCD frame made in Ostrowik observatory. 
D.B./R.K. Start and D.B./R.K End correspond to data obtained by 
David Boyd and Robert Koff respectively. 
"Dur." Gives the total duration of runs from all telescopes 
excluding the gaps.}
\centering
\smallskip
{\small
\begin{tabular}{l c c c c c c c c c} \hline \hline
Date in & O. Start & O. End   &D.B. Start  & D.B. End   & R.K. Start&R.K. End & Dur. & No. of & $<V>$\\
2006    & 2453000+ & 2453000+ & 2453000+ & 2453000+ & 2453000+ & 2453000+ & [hr] & points & [mag] \\
\hline
Jul 17/18 & 934.35741 & 934.48498 & - & - & - & - & 3.07 & 96  & 14.85 \\
Jul 18/19 & 935.34981 & 935.46236 & 935.4153 & 935.5366 & - & - & 4.48 & 430 & 14.87 \\
Jul 19/20 & 936.34761 & 936.45140 & - & - & - & - & 2.49 & 90  & 14.97 \\
Jul 20/21 & 937.34381 & 937.53570 & 937.4028 & 937.5556 & - & - & 5.08 & 524 & 15.07 \\
Jul 21/22 & 938.33272 & 938.35473 & 938.4610 & 938.5400 & 938.6425 & 938.9360 & 9.47 & 486 & 15.23 \\
Jul 22/23 & 939.34842 & 939.46027 & 939.4799 & 939.5819 & 939.6373 & 939.9487 & 12.06 & 717 & 15.31 \\
Jul 23/24 & 940.33218 & 940.46082 & 940.4018 & 940.5223 & - & - & 4.56 & 560 & 15.41 \\
Jul 24/25 & - & - & 941.4139 & 941.5463 & - & - & 3.18 & 426 & 15.54 \\
Jul 25/26 & 942.33893 & 942.46321 & - & - & - & - & 2.98 & 107 & 15.62 \\
Jul 26/27 & 943.33482 & 943.39914 & - & - & 943.6480 & 943.8573 & 6.57 & 239  & 15.84 \\
Jul 27/28 & 944.32455 & 944.44318 & - & - & 944.6359 & 944.9476 & 10.33 & 398 & 16.59 \\
Jul 28/29 & 945.33554 & 945.44203 & - & - & - & - & 2.56 & 67 & 17.36 \\
Jul 30/31 & 947.34226 & 947.43608 & - & - & - & - & 2.25 & 32 &  18.14 \\
Aug 2/3   & 950.32019 & 950.44469 & - & - & - & - & 2.99 & 87 & 17.63 \\
\hline
\label{table1}
\end{tabular}}
\end{table}

\section{Global light curve}

Figure 1 presents the photometric behaviour of V419 Lyr in 2006 July and
August. Dots and open circles correspond to our CCD observations
and visual estimates of AAVSO observers, respectively.

\vspace{7.5cm}

\includegraphics{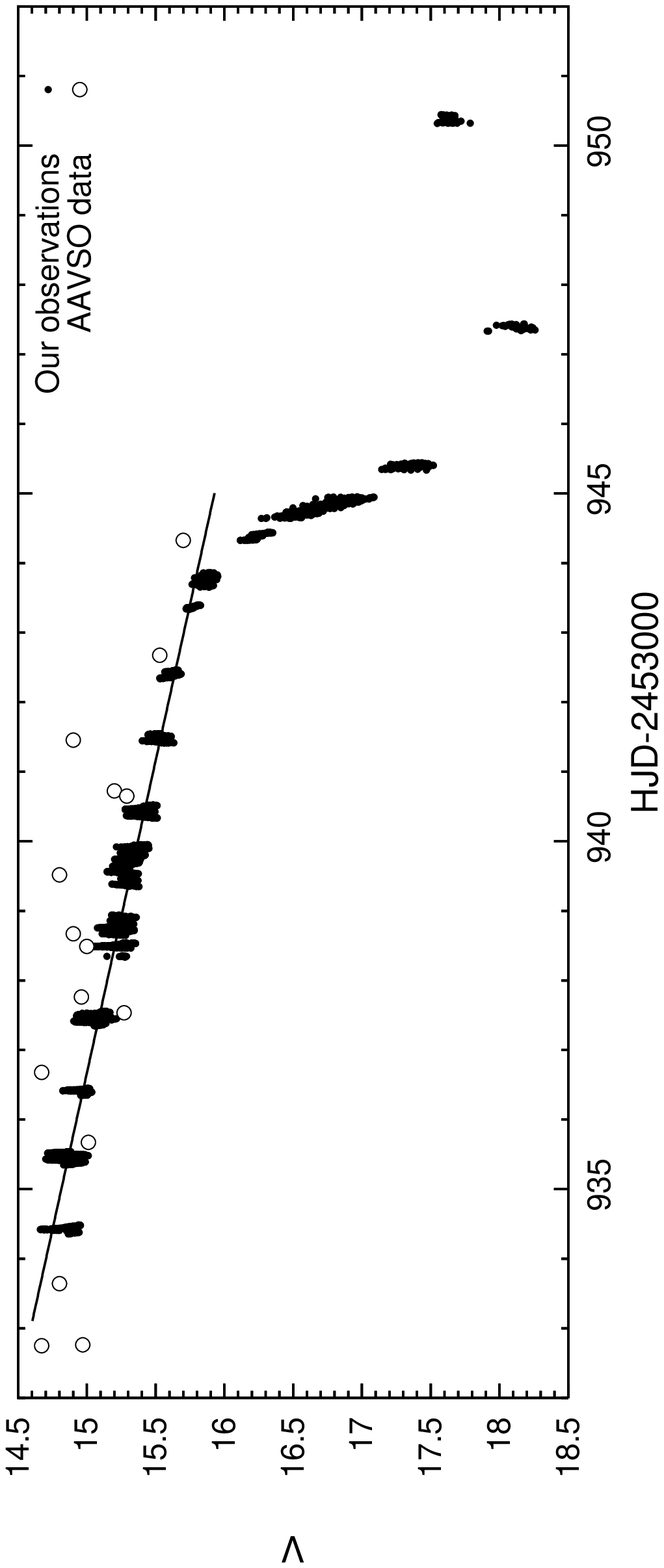}

\begin{figure}[h]
\caption{\sf Global light curve of 2006 superoutburst of V419 Lyr.
Dots and open circles denote our observations and AAVSO estimates,
respectively. Solid line is a least squares linear fit to the plateau
phase of the superoutburst.}
\end{figure}

The shape of the light curve corresponds to the standard picture of a
superoutburst. First, the star rose rapidly from its quiescent
level to a peak magnitude of around 14.5 mag. Due to the lack of
observations, the exact time of the rise is unknown. V419 Lyr was seen
by AAVSO observers in a bright state on July 16. Two nights later, during
our first run, the star was still at the same brightness and
showing fully developed superhumps with a peak-to-peak amplitude of 0.3 mag - a
feature characteristic of the beginning of a superoutburst. Thus we conclude
that the July 2006 superoutburst started around July 15.

\vspace{17cm}

\includegraphics{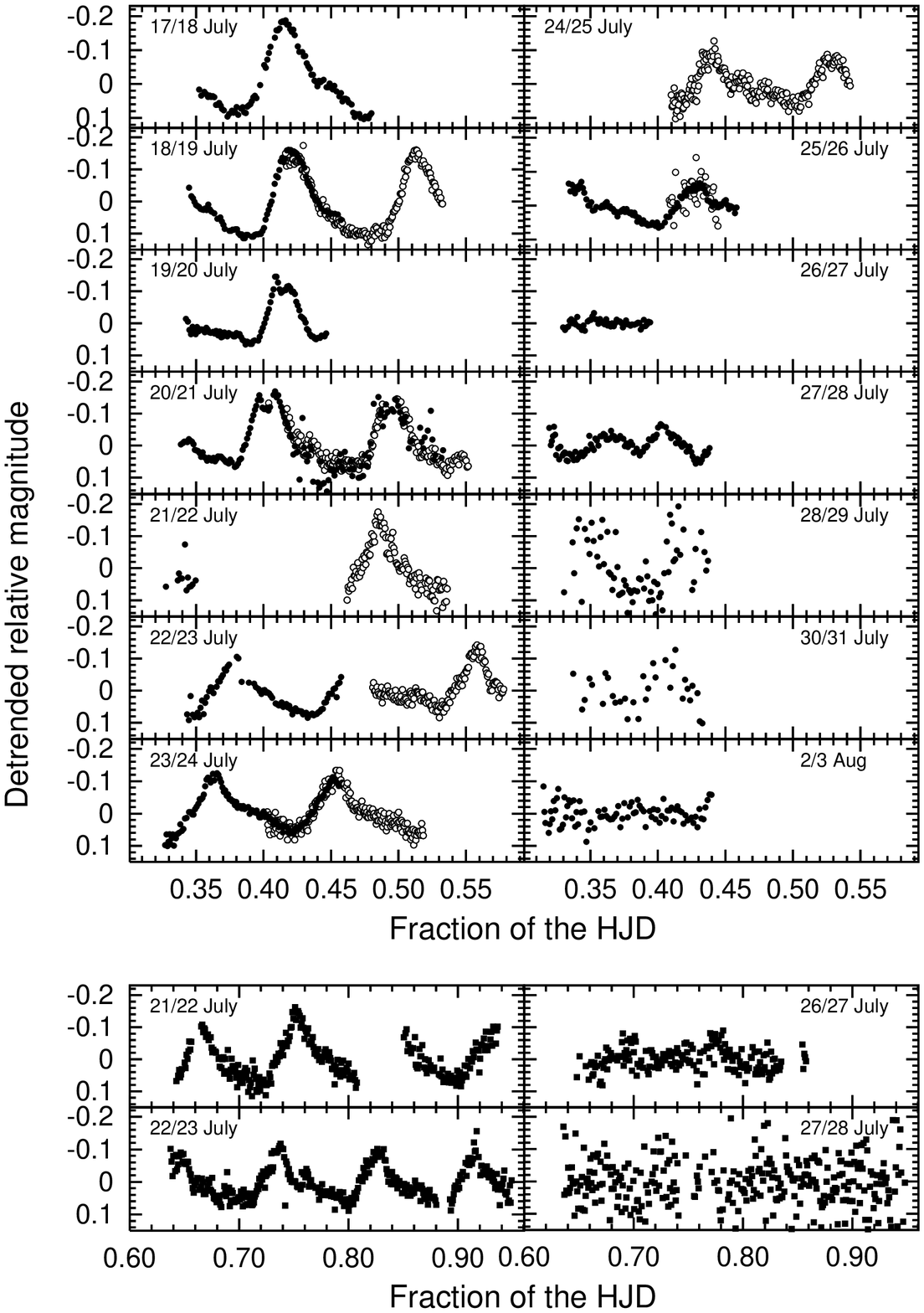}

\begin{figure}[h]
\caption{\sf Superhumps observed each night during the 2006 July
superoutburst of V419 Lyr.}
\end{figure}

After reaching peak magnitude, V419 Lyr entered the plateau stage
lasting about 11 days with an average decline rate of about 0.1 mag d$^{-1}$.
The plateau stage ended rapidly on July 27 and during the next two days we
observed the final decline stage with a change of brightness of about 1 mag
d$^{-1}$. On July 30 the star was again in quiescence showing a mean
brightness of 18.1 mag. Thus the entire superoutburst lasted about 15
days.

The linear fit to the plateau stage in Figure 1 is helpful in showing
there was no rebrightening as has been observed in some SU UMa systems
(Kato et al. 2003a).

\section{Superhumps}

Figure 2 shows the light curves of V419 Lyr during thirteen individual
nights. Filled circles correspond to the Ostrowik Observatory data,
while open circles and squares represent D.B. and R.K. measurements respectively.

The magnitudes have been transformed to a common $V$ system and detrended
for purposes of Fourier power spectrum analysis. The superhumps are
clearly visible and have an initial amplitude of 0.3 mag.  Later the
superhump amplitude gradually decreases and on August 2/3 is
practically indistinguishable from noise.

\subsection{ANOVA statistics}

As we noted earlier, all light curves of V419 Lyr in superoutburst were
detrended by removing a fit based on a first or second order
polynomial. Then we analyzed them using {\sc anova} statistics
(Schwarzenberg-Czerny 1996). The resulting periodogram is shown in Figure
3. It shows a very clear and dominant peak at frequency $f_0 = 11.107 \pm
0.010$ c/d, which we interpret as due to superhumps and corresponds to the period $P_{\rm sh} = 0.09033(81)$ days.

\vspace{7.9cm}

\includegraphics{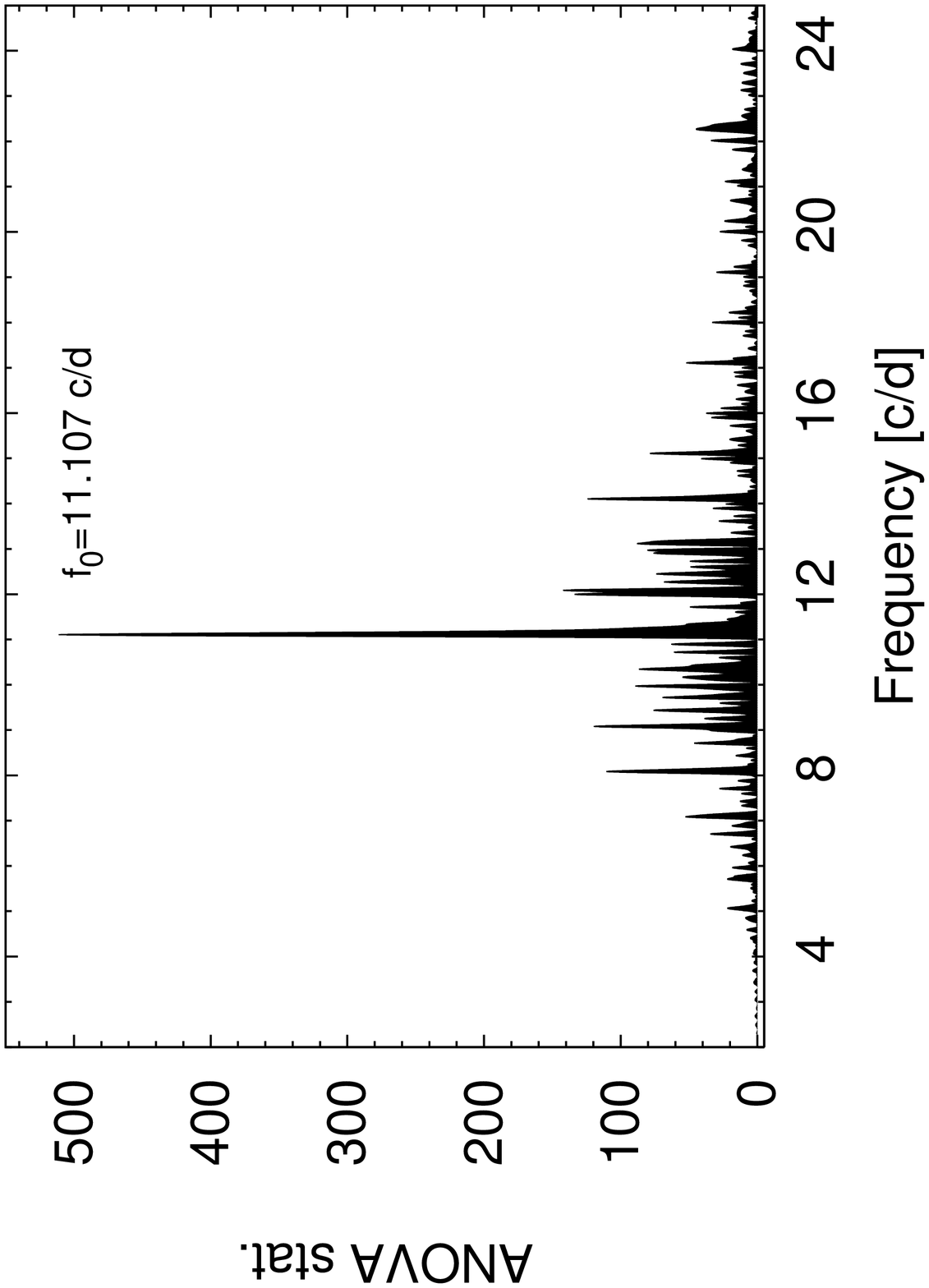}

\begin{figure}[h]
\caption{\sf ANOVA power spectrum computed for data from nights July 17/18
- July 27/28.}
\end{figure}

The spectrum has almost no 1-day aliases due to good coverage and the
use of data from three sites - two from Europe and one from the United
States. A small peak is also observed around 22 c/d, the first harmonic
of the main frequency. We then prewhitened the light curve of V419 Lyr
during the superoutburst with the main period and its first harmonic.
The power spectrum of the resulting light curve shows no other
periodicities.

\subsection{The O-C analysis}

To check the stability of the superhump period and to better determine its
value, we constructed an $O-C$ diagram. Because the maxima in our case
were almost always clearly visible and easier to measure
than minima, we decided to use the timings of the former.  Finally, we
were able to determine 27 times of maxima which are listed in Table 2
together with associated errors, cycle numbers $E$, and $O-C$ values.

\begin{table}[!h]
\caption{Times of maxima observed in the light curve of V419 Lyr during 
its 2006 superoutburst}
\smallskip
\centering
\begin{tabular}{r c c r} \hline \hline
Cycle & HJD-2 453 000 & Error & O-C \\ \hline
0  & 934.420  & 0.0015 & -0.1408\\
11 & 935.425  & 0.0014 & 0.0281\\
12 & 935.517  & 0.0016 & 0.0505\\
22 & 936.420  & 0.0016 & 0.0859\\
33 & 937.408  & 0.0014 & 0.0660\\
34 & 937.497  & 0.0017 & 0.0551\\
45 & 938.490  & 0.0031 & 0.0907\\
47 & 938.669  & 0.0022 & 0.0800\\
48 & 938.757  & 0.0025 & 0.0580\\
55 & 939.383  & 0.0044 & 0.0149\\
57 & 939.562  & 0.0031 & 0.0042\\
58 & 939.650  & 0.0034 &-0.0177\\
59 & 939.741  & 0.0029 &-0.0064\\
60 & 939.831  & 0.0033 &-0.0062\\
61 & 939.920  & 0.0033 &-0.0171\\
66 & 940.369 & 0.0017 & -0.0272\\
67 & 940.458 & 0.0025 & -0.0381\\
78 & 941.443 & 0.0027 & -0.0914\\
79 & 941.533 & 0.0030 & -0.0912\\
88 & 942.344 & 0.0015 & -0.0782\\
89 & 942.432 & 0.0033 & -0.1002\\
103& 943.696 & 0.0047 & -0.0530\\
104& 943.783 & 0.0044 & -0.0861\\
111& 944.407 & 0.0022 & -0.1513\\
121& 945.335 & 0.0056 & 0.1663\\
122& 945.425 & 0.0056 & 0.1643\\
144& 947.411 & 0.0051 & 0.2322\\
\hline
\end{tabular}
\end{table}
\bigskip

A least-squares linear fit to the data taken during the plateau phase of
superoutburst gives the following ephemeris for the maxima:

\begin{equation}
\textrm{HJD}_\textrm{max} = 2453934.4326(41)+0.089983(58) \cdot E
\end{equation}

The above equation indicates that the mean superhump period was $P_{\rm sh} =
0.089983(58)$ days, which agrees within errors with the determination based
on {\sc anova} statistics. Combining these two measurements
gives us our final estimate of the mean superhump period of V419 Lyr
during its 2006 July superoutburst which is $P_{\rm
sh}=0.089985(58)$ days ($129.58\pm0.08$ min).

The $O-C$ values computed according to the ephemeris (1) are listed in
Table 2 and also shown in Figure 4. It is clear that V419 Lyr, during its
2006 superoutburst, showed clear changes of superhump period. In the
cycle range $0-70$ the period was quickly decreasing. A second-order
polynomial fit to $\textrm{E} ~vs.~ \textrm{HJD}_{\textrm{max}}$ dependence
in this range corresponds to the solid line in the bottom panel of Figure 4 and
is expressed by the following ephemeris:

\begin{equation}
\textrm{HJD}_\textrm{max} =  2453934.42496(117) + 0.0908080(75) \cdot E -
1.125(97)\cdot 10^{-5} \cdot E^2
\end{equation}

This equation indicates that the period derivative has the large value
$\dot{P}/P_{sh}=-24.8(2.2)\times10^{-5}$. This is the second largest
negative value detected in SU UMa stars. Figure 5, taken from Kato et al.
(2003b) and Olech et al. (2003), shows the position of V419 Lyr on the
$\dot{P}/P_{sh} ~vs~ P_{\rm sh}$ diagram. Only KK Tel had a
faster period decrease during its 2002 June superoutburst (Kato et
al. 2003b). It is interesting that both V419 Lyr and KK Tel are long
superhump period dwarf novae with orbital periods very close or even
within the period gap.

\vspace{8.7cm}

\includegraphics{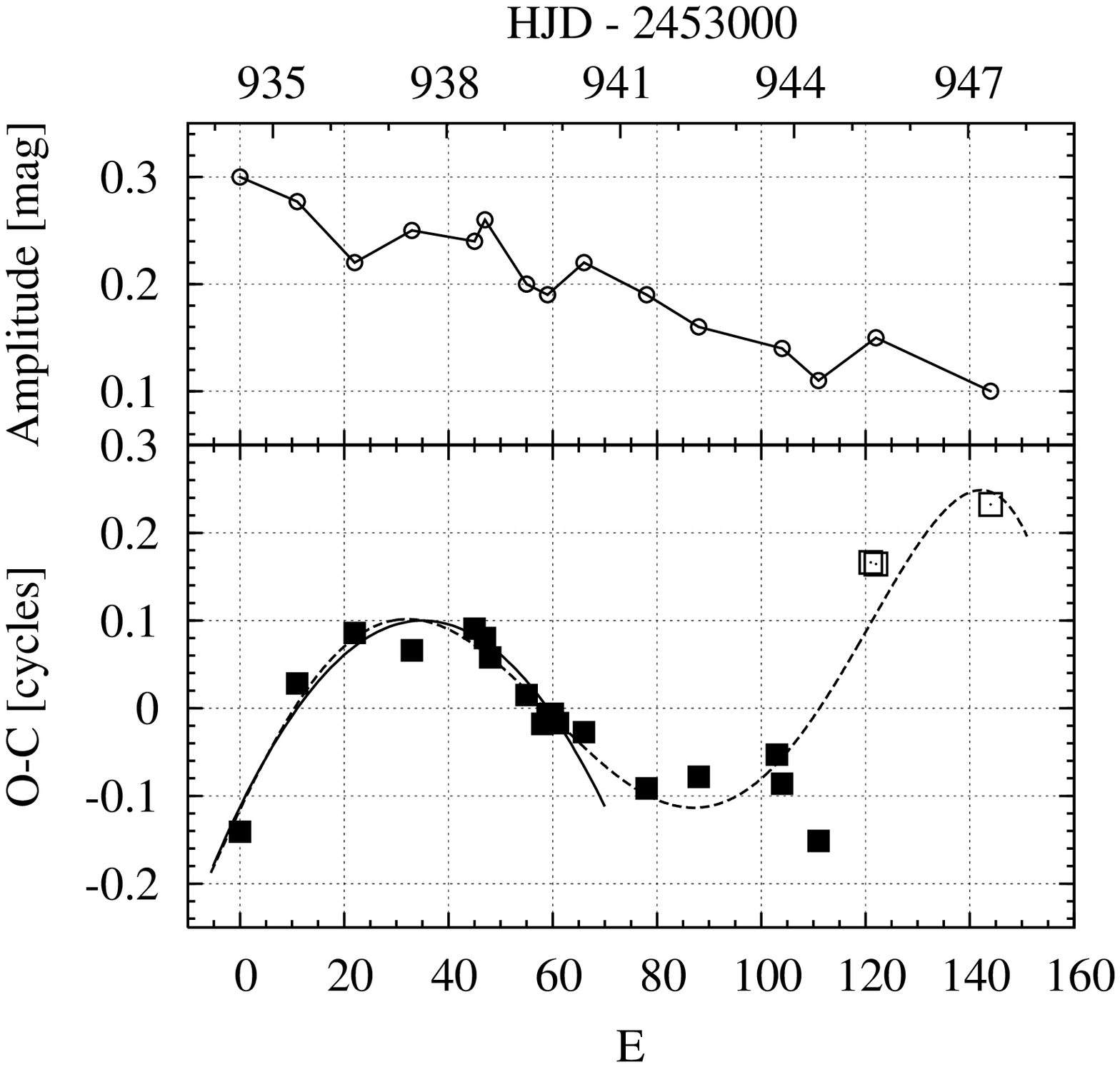}

\begin{figure}[h]
\caption{\sf Upper panel: Evolution of the amplitude of superhumps
observed in the 2006 July superoutburst of V419 Lyr. Lower panel: $O-C$ diagram
for times of superhump maxima. Solid line corresponds to the quadratic
ephemeris (2). Black and open squares denote normal and late superhump maxima
respectively. Dashed line corresponds to the 5th order polynomial fit. Size
of the squares is, on average, of the size of error bars.}
\end{figure}

\vspace{7.9cm}

\includegraphics{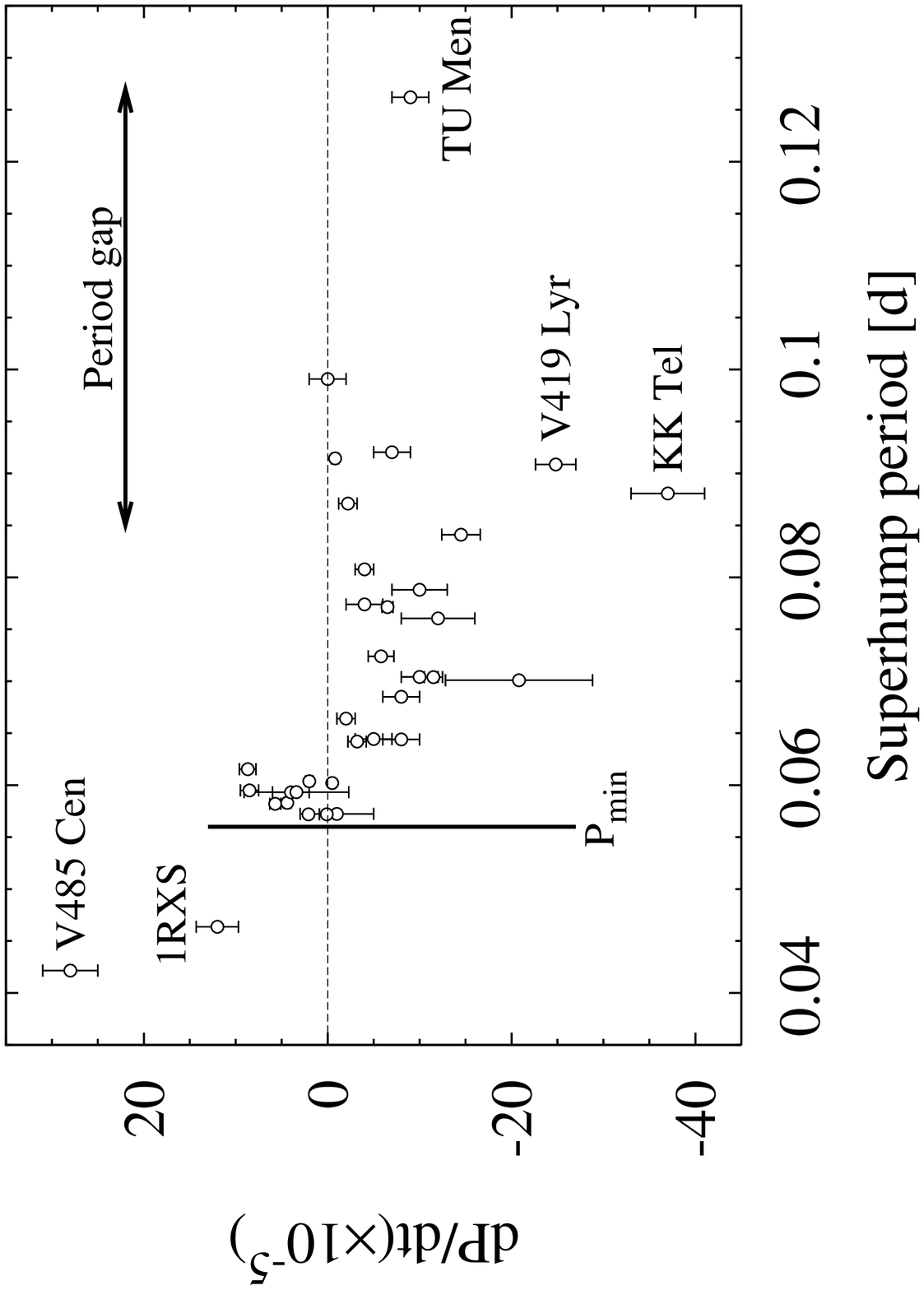}

\begin{figure}[h]
\caption{\sf $\dot P/P_{sh}$ versus $P_{sh}$ for SU UMa-type dwarf novae.
The figure is taken from Kato et al. (2003b) and Olech et al. (2003). 
$P_{min}$ denotes the boundary for a hydrogen rich secondary. The range of 
the period gap is also plotted. 1RXS is an abbreviation for 1RXS J232953.9+062814.}
\end{figure}

It is worth commenting on the behaviour of the superhump period after
cycle number $E=70$. At that moment, the star was still in the plateau
phase of the superoutburst, three days before entering the final decline,
and superhumps were still clearly visible in the light curve. The $O-C$
values for cycle numbers from 70 to 111 can be roughly fitted with a
straight line, which indicates that the period decrease had stopped and
its value stabilized at $P_{\rm sh} = 0.08983(8)$ days.

The noisy maxima with cycle numbers above 120 are shifted by $\sim 0.4$
cycle with respect to the superhump maxima from earlier nights. This
indicates that they may be connected with late superhumps or
even with the orbital wave.

It is interesting that our $O-C$ diagram could be also interpreted in
diffrent way than "common superhump followed by transition to late
superhump" scenario. Fitting a 5th order polynomial to the moments of
maxima appears to work with the data about as well as a quadratic
followed by a linear trend as described previously. We are not arguing
in favour of this interpretation, simply saying that there might be
other interpretations of the data.

\subsection{Amplitude and shape}

The upper panel of Figure 4 shows the evolution of the amplitude of
superhumps through the entire superoutburst. In a typical SU UMa star,
fully developed superhumps have an amplitude of about 0.3 mag and a
characteristic tooth shape. Interestingly, these properties seem to be
completely independent of the inclination of the orbit of the binary. As
the outburst progresses, the amplitude gradually decreases and the
profile of the humps changes. A few days after maximum, so called
secondary humps or interpulses become visible. In the beginning they are
small but with time they may become as high as the main maxima - most
probably evolving towards late superhumps.

V419 Lyr is an interesting case because it seems not to follow this
scenario. On the nights of July 19/20 and 20/21 it clearly showed double
maxima. However, these quickly disappeared. During subsequent nights
there was hardly a trace of secondary humps. Very weak humps could be
seen only on July 23/24 and 25/26 and quite strong ones on July 27/28.

We were curious about the reason for such behaviour. The only property
which strongly differs in V419 Lyr from typical SU UMa stars is its long
superhump/orbital period placing it within the period gap. We therefore
reviewed the literature to investigate the occurrence of secondary humps
among long period systems (i.e. with superhump period $>0.08$ days). The
summary of our review is given in Table 3.

\begin{center}
\begin{table}[!h]
\caption{Occurrence of secondary humps in long period superhumpers}
\smallskip
\centering
\begin{tabular}{l l c c c c c} \hline \hline
Star        &    $P_{\rm sh}$ [days]  &       Clear   &        Poorly    &      Invisible   &    Ref  \\ \hline
HS Vir      &    0.08077  &       -       &        +         &      -           & 1,2      \\
V359 Cen    &    0.08092  &       -       &        +         &      -           & 3,4      \\
V660 Her    &    0.081    &       -       &        -         &      +           & 5        \\
V503 Cyg    &    0.08101  &       +       &        -         &      -           & 6        \\
BR Lup      &    0.08220  &       +       &        -         &      -           & 7,8       \\
V877 Ara    &    0.08411  &       +       &        -         &      -           & 9           \\
AB Nor      &    0.08438  &       -       &        +         &      -           & 10          \\
V369 Peg    &    0.08484  &       -       &        -         &      +           & 11           \\
HV Aur      &    0.08559  &       +       &        -         &      -           & 12           \\
EF Peg      &    0.08705  &       -       &        +         &      -           & 13,14         \\
TY PsA      &    0.08765  &       -       &        +         &      -          &  15,16         \\
BF Ara      &    0.08797  &       -       &        -         &      +           & 17            \\
KK Tel      &    0.08803  &       -       &        -         &      +           & 18            \\
DV UMa      &    0.08869  &       -       &        +         &      +           & 19,20           \\
V419 Lyr    &    0.0901   &       -       &        +         &      -           & this study,21   \\
UV Gem      &    0.0902   &       -       &        +         &      -           & 22              \\
V344 Lyr    &    0.09145  &       -       &        -         &      +           & 23              \\
YZ Cnc      &    0.09204  &       -       &        +         &      -           & 24,25           \\
GX Cas      &    0.09297  &       -       &        -         &      +           & 26              \\
V725 Aql    &    0.09909  &       -       &        +         &      -           & 27              \\
MN Dra      &    0.1055   &       -       &        -         &      +           & 28,29           \\
\hline
\multicolumn{6}{l}{\small 1. Kato et al. (1998), 2. Kato et al. (2001), 3. Woudt \& Warner (2001), 4. Kato et al. (2002),} \\
\multicolumn{6}{l}{\small 5. Olech et al.(2005), 6. Harvey et al.(1995), 7. Mennickent \& Sterken (1998), 8. O'Donoghue (1987),}\\
\multicolumn{6}{l}{\small 9. Kato et al. (2003c), 10. Kato et al. (2004b), 11. Kato \& Uemura (2001), 12. Nogami et al. (1995),}\\
\multicolumn{6}{l}{\small 13. Kato (2002), 14. Howell et al. (1993), 15. Barwig et al. (1982), 16. Warner et al. (1989),} \\
\multicolumn{6}{l}{\small 17. Kato et al. (2003c), 18. Kato et al. (2003b), 19. Nogami et al. (2001), 20. Feline et al. (2004),}\\
\multicolumn{6}{l}{\small 21. Nogami et al. (1998), 22. Kato \& Uemura (2001), 23. Kato (1993), 24. Patterson (1979),}\\
\multicolumn{6}{l}{\small 25. Kato (2001), 26. Nogami et al.(1998), 27. Uemura et al. (2001), 28. Antipin \& Pavlenko (2002),}\\
\multicolumn{6}{l}{\small 29. Nogami et al. (2003)}\\
\end{tabular}
\end{table}
\end{center}

Among 21 reviewed stars only four show clear secondary humps. The rest
of them show no interpulses at all or only a weak trace of them.

\section{Conclusions}

The orbital period of V419 Lyr is unknown. However it is possible to
estimate its value using the relation in Stolz and Schoembs (1984) connecting the
period excess $\epsilon$ defined as $P_{\rm sh}/P_{\rm orb}-1$ with the
orbital period of the binary. This empirical relation is as follows:

\begin{equation}
\epsilon = 0.858(11) \cdot P_{\rm orb} - 0.0282(2)
\end{equation}

Using the definition of $\epsilon$ and knowing $P_{\rm sh}$ for V419 Lyr, we
were able to estimate the orbital period as $P_{\rm
orb}\approx 0.086$ days. This is slightly longer than two hours which
indicates that V419 Lyr is a dwarf nova in the period gap.

Many characteristics of V419 Lyr are typical of SU UMa stars. It goes
into superoutburst every year or so, the eruption lasts about two weeks
and has an amplitude of $\sim 3.5$ mag. Superhumps appear shortly after the
beginning of the superoutburst and have a maximum amplitude of 0.3 mag,
which decreases to 0.1 mag at the end of the outburst. In addition to its long
orbital period, V419 Lyr is unusual in two other properties. Its
superhump period derivative has one of the largest negative values known
and it shows only a weak trace of secondary humps in the final stages
of the superoutburst.

\bigskip \noindent {\bf Acknowledgments.} ~We acknowledge generous
allocation of  the Warsaw Observatory 0.6-m telescope time. This work
used the online service of the VSNET and AAVSO. We would like to thank 
Prof. J\'ozef Smak for fruitful discussions.

\end{document}